\documentstyle[twocolumn,aps,epsf,amsbsy]{revtex}

\begin{document}

\title{Comment on ``Simple Measure for Complexity''}
\author{James P. Crutchfield,
David P. Feldman,\thanks{Permanent address: College of the Atlantic,
105 Eden Street, Bar Harbor, Maine 04609.}
and Cosma Rohilla Shalizi\thanks{Permanent address: Physics Department,
University of Wisconsin, Madison, WI 53706.}
}
\address{Santa Fe Institute, 1399 Hyde Park Road, Santa Fe, NM 87501\\
Electronic address: \{chaos,dpf,shalizi\}@santafe.edu}

\date{\today}
\maketitle

\begin{abstract}
We critique the measure of complexity introduced by Shiner, Davison, and
Landsberg in Ref.~ \cite{Shin99a}. In particular, we point out that it is
over-universal, in the sense that it has the same dependence on disorder for
structurally distinct systems. We then give counterexamples to the claim that
complexity is synonymous with being out of equilibrium: equilibrium systems
can be structurally complex and nonequilibrium systems can be structurally
simple. We also correct a misinterpretation of a result given by two of the
present authors in Ref.~ \cite{Crut97a}.
\end{abstract}

\begin{center}
Santa Fe Institute Working Paper 99-06-040
\end{center}


In Ref.~\cite{Shin99a}, Shiner, Davison, and Landsberg introduce a
two-parameter family $\Gamma_{\alpha\beta}$ of complexity measures:
\begin{equation}
    {\Gamma}_{\alpha\beta} \, \equiv \, \Delta ^\alpha ( 1 -
    \Delta)^\beta \;, 
\label{C_SDL}
\end{equation}
where
\begin{equation}
    \Delta \, \equiv \, \frac{S}{S_{\rm max}} \;.
\end{equation} 
The quantity $\Delta$ is called the ``disorder'', $S$ is the
Boltzmann-Gibbs-Shannon entropy of the system, and ${S}_{ \rm max}$ its maximum
possible entropy---taken to be equal to the equilibrium thermodynamic
entropy. For $\alpha, \beta > 0$, $\Gamma_{\alpha\beta}$ satisfies the widely
accepted ``one-hump'' criterion for statistical complexity measures---the
requirement that any such measure be small for both highly ordered and highly
disordered systems \cite{Gras86,Lind88b,Crut89,Gell96,Badi97}. The approach to
complexity measures taken by Shiner, Davison, and Landsberg \cite{Shin99a} is
similar to that of L\`{o}pez-Ruiz, Mancini, and Calbet \cite{Lope95a}.  In both
Refs.~\cite{Shin99a} and \cite{Lope95a} the authors obtain a measure of
complexity satisfying the one-hump criterion by multiplying a measure of
``order'' by a measure of ``disorder''.

We welcome this addition to the literature on complexity measures and are
pleased to see a variety of complexity measures compared and examined
critically.  However, there are several aspects of Ref.~\cite{Shin99a} upon
which we would like to comment.

First, despite satisfying the one-hump criterion, it is not clear that
$\Gamma_{\alpha\beta}$ is a measure of {\it complexity}.
$\Gamma_{\alpha\beta}$ is a quadratic function of a measure of
distance from thermodynamic equilibrium, as the authors note on p.~ 1461. This
has three consequences:

\begin{enumerate}

\item As pointed out in Ref.~\cite{Feld97a}, this type of complexity measure is
over-universal in the sense that it has the same dependence on disorder for
structurally distinct systems.  Eq.~(\ref{C_SDL}) makes it clear that, despite
the claims of Shiner et al. to the contrary, all systems with the same disorder
$\Delta$ have the same $\Gamma_{\alpha\beta}$.

\item Since $S_{\rm max}$ is taken to be the equilibrium entropy of the system,
$\Gamma_{\alpha\beta}$ vanishes for {\it all} equilibrium systems:
``\,`Complexity' vanishes ... if the system is at equilibrium''
\cite[p. 1461]{Shin99a}. Due to this $\Gamma_{\alpha\beta}$ does not
distinguish between two-dimensional Ising systems at low temperature, high
temperature, or the critical temperature.  All of these systems are at
equilibrium and hence have vanishing $\Gamma_{\alpha\beta}$. However, they
display strikingly different {\it degrees} of structure and organization.  Nor
does $\Gamma_{\alpha\beta}$ distinguish between the many different {\it kinds}
of organization observed in equilibrium \cite{Chai95a}---between, say, ideal
gases, the long-range ferromagnetic order of low-temperature Ising systems, the
orientational and spatial order of the many different liquid crystal phases
\cite{Coll97}, and the intricate structures formed by amphiphilic systems
\cite{Gomp94}.  All of these systems are in equilibrium, but they (presumably)
have very different complexities.

\item We have just seen that equilibrium should not be taken to
indicate an absence of complexity.  Conversely, not all systems out of
equilibrium are complex.  For example, consider a paramagnet, a
collection of two-state spins that are not coupled.  If this system is
pumped so that it's out of equilibrium, a larger percentage of the
spins will be in their higher energy states.  Nevertheless, there is
still no spatial structure or ordering in the system; the spins are
still completely uncorrelated.  However, the complexity measure of
Shiner et al. will be nonzero for this very simple system.  While 
$\Gamma_{\alpha\beta}$ vanishes for systems at ``maximal distance from
equilibrium'' Ref.~\cite[p. 1461]{Shin99a}, all other systems
displaced from equilibrium have non-vanishing complexity by virtue of
the $1 - \Delta$ term in Eq.~(\ref{C_SDL}).  It does not seem
reasonable to us to require that {\em any}\/ system partially out of
equilibrium have positive complexity.

\end{enumerate}

In summary, then, we argue that whether or not a system is in
equilibrium in and of itself says little about the system's structure,
pattern, organization, or symmetries. Equilibrium systems can be
complex, nonequilibrium systems can be simple, and vice versa. Since
$\Gamma_{\alpha \beta}$ is defined in terms of a ``distance from
equilibrium'' $1 - \Delta$, we feel that it cannot capture structural
complexity. 

Second, we are confused by Ref.~\cite{Shin99a}'s calculation of $\Gamma_{11}$
for equilibrium Ising systems on p.~1462.  If the system is at equilibrium,
then the disequilibrium term $1 - \Delta$ should vanish, leading to a vanishing
$\Gamma_{11}$.  Perhaps the authors are using a uniform distribution rather
than the thermodynamic equilibrium distribution in their calculation of $S_{\rm
max}$.

Third, Ref.~\cite{Shin99a} appears to have misinterpreted our earlier work on
the statistical complexity of one-dimensional spin systems
\cite{Crut97a,Feld97b}. On p.~1462, Ref.~\cite{Shin99a} identifies the
statistical complexity $C_\mu$ \cite{Crut89,Crut92c} with zero-coupling ($J=0$)
disorder $\Delta$. At a minimum, this interpretation is not consistent
dimensionally, since $C_\mu$ has the units of entropy (bits), while $\Delta$ is
a dimensionless ratio.  More crucially, however, Ref.~\cite{Shin99a} conflates
the {\it definition} of $C_\mu$, which does not make $C_\mu$ a function solely
of the system's entropy, with {\it a particular equation} for $C_\mu$ (Eq.~(8)
of Ref.~\cite{Crut97a}) correct within a strictly delimited range of validity
\cite{Crut97a,Feld97b}.  Further, Ref.~\cite{Shin99a} draws an inaccurate
conclusion based on that equation. For nearest-neighbor Ising systems
Refs. \cite{Crut97a} and \cite{Feld97b} show that $C_\mu = H(1)$, the entropy
of spin blocks of length one. Contrary to the statement in Ref.~\cite{Shin99a},
$H(1)$ is not the same as the entropy of noninteracting spins---i.e., of
paramagnetic spin systems, those with $J=0$.

Finally, Ref.~\cite{Shin99a} states that thermodynamic depth \cite{Page88a}
belongs to the family of complexity measures that are single-humped functions
of disorder.  However, two of us recently pointed out that thermodynamic
depth is an increasing function of disorder \cite{Crut98d}.

In summary, we have argued here and elsewhere \cite{Feld97b,Crut92c} that a
useful role for statistical complexity measures is to capture the
structures---patterns, organization, regularities, symmetries---intrinsic to a
process.  Ref.~\cite{Feld97a} emphasizes that defining such measures solely in
terms of the one-hump criterion---say, by multiplying ``disorder'' by ``one
minus disorder''---is insufficient to this task.  Introducing an arbitrary
parameterization of this product---e.g. via $\alpha$ and $\beta$ in
Eq.~(\ref{C_SDL})---does not help the situation.  A statistical complexity
measure that is a function only of disorder is not adequate to measure
structural complexity, since it is unable to distinguish between structurally
distinct configurations with the same disorder.

This work was supported at the Santa Fe Institute under the Computation,
Dynamics and Inference Program via ONR grant N00014-95-1-0975 and by Sandia
National Laboratory.  We thank an anonymous referee for several
helpful comments.

\bibliographystyle{unsrt}
\bibliography{spin}

\end{document}